\documentclass[english,a4paper,12pt]{article}
\usepackage[T2A]{fontenc}
\usepackage[cp1251]{inputenc}
\usepackage{amsmath}
\usepackage{graphicx}
\usepackage{amssymb}
\usepackage{epsf}
\usepackage{pazh_eng}
\usepackage[labelsep=period,hang]{caption} 

\tightenlines

\newcommand{\NMS}[1]{\mbox{$#1\,$M$_{\textstyle\odot}$}}
\newcommand{\nucmu}{\mbox{$m_{\mathrm u}$}}
\newcommand{\WID}[2]{#1_{\mathrm{#2}}}
\newcommand{\DER}[2]{\frac{\partial #1}{\partial #2}}
\newcommand{\DERS}[3]{\left(\frac{\partial #1}{\partial #2}\right)_{\!\!#3}}

\sloppypar

\begin{document}
\baselineskip 21pt


\title{\bf Special Point on the Mass–Radius Diagram of Hybrid Stars}

\author{\bf \hspace{-1.3cm}\ \
A.V. Yudin\affilmark{1,2*}, T.L. Razinkova\affilmark{1}, D.K.
Nadyozhin\affilmark{1,3}, A. D. Dolgov\affilmark{1,2,4}}

\affil{ {\it Institute for Theoretical and Experimental Physics,
ul. Bolshaya Cheremushkinskaya 25, Moscow, 117259
Russia}$^1$\\
{\it Novosibirsk State University, ul. Pirogova 2, Novosibirsk, 630090 Russia}$^2$\\
{\it “Kurchatov Institute” National Research Center, pl. Kurchatova 1, Moscow, 123182 Russia}$^3$\\
{\it University of Ferrara and INFN, Ferrara, Italy}$^4$}

\vspace{2mm}

\sloppypar \vspace{2mm} \noindent
  An analytical study that explains the existence of a very small region on the mass–radius
   $(M{-}R)$ diagram of hybrid stars where all of the lines representing the sequences of models with different
constant values of the bag constant $B$ intersect is presented.
This circumstance is shown to be a consequence of the linear
dependence of pressure on energy density in the quark cores of
hybrid stars.



\vfill
\noindent\rule{8cm}{1pt}\\
{$^*$ e-mail $<$yudin@itep.ru$>$}

\clearpage

\section*{INTRODUCTION}
In recent years, hydrostatically equilibriummodels of superdense
hybrid stars that consist of a quark core and an outer crust of
nuclear matter have been widely discussed in scientific literature
(see, e.g., the book by Haensel et al. (2007) and references
therein). The properties of hybrid stars are of great importance
for explaining the supernova explosionmechanism in the simplest
case where there are no magnetic field and rotation. This is
because the phase transition to quark matter that arises at the
boundary between the core of a hybrid star and its crust can be
responsible for the development of hydrodynamic instability ending
with a supernova explosion (see Yudin et al. (2013) and references
therein).

The published models of hybrid stars show a surprising
peculiarity. On the mass–radius $(M{-}R)$ diagram, all of the
lines representing the sequences of models with different constant
values of the bag constant $B$ intersect in a very small region
that we arbitrarily call a “point” here. As far as we know, there
is no discussion of this fact in the literature. In this paper, we
present an analytical study that hopefully remedies this
deficiency.

\section*{FORMULATION OF THE PROBLEM}
 To construct the stellar models, we use an equation of
state (EOS) with the phase transition to quark matter at high
densities (for more details, see Yudin et al. 2013). An
approximation of the EOS from Douchin and Haensel (2001) is
applied for the low density component of the matter. The quark
component is described by the simplest version of the bag model in
which the relation between pressure P and total energy per unit
volume $\epsilon$ is linear:
\begin{equation}
P=\frac{1}{3}(\epsilon-4B),\label{P-E-B}
\end{equation}
where $B$ is the quark bag constant. This approximation is widely
used in modelling the properties of quark matter and is a special
case of the group of linear EOSs: $P=\alpha(\epsilon-\epsilon_0)$,
where the dimensionless constant $0\leq\alpha\leq 1$ means the
square of the speed of sound measured in units of the speed of
light, $\alpha=(\WID{c}{s}/c)^2$. The bag constant $B$ is a free
model parameter and, in the simplest case, is uniquely related to
the density at which the phase transition begins. The transition
itself is an ordinary first-order phase transition with
$\left(\frac{\partial P}{\partial\rho}\right)_T=0$ is a free model
parameter and, in the simplest case, is uniquely related to the
density at which the phase transition begins. The transition
itself is an ordinary first-order phase transition with
$\left(\frac{\partial P}{\partial\rho}\right)_T>0$, in the phase
coexistence region, while the region itself expands. However, the
question about allowance for the interaction between the phases in
a mixed state arises for such a description. The self-consistent
calculation of these effects is rather complex. In this case, for
example, Maruyama et al. (2007) showed that such allowance makes
the resulting phase transition much more similar in properties to
the simple Maxwellian distribution. Therefore, we conclude that
the Maxwellian approach is a good approximation for our goal.

\begin{figure}[htb]
\begin{center}
\epsfxsize=14cm \hspace{-0cm}\epsffile{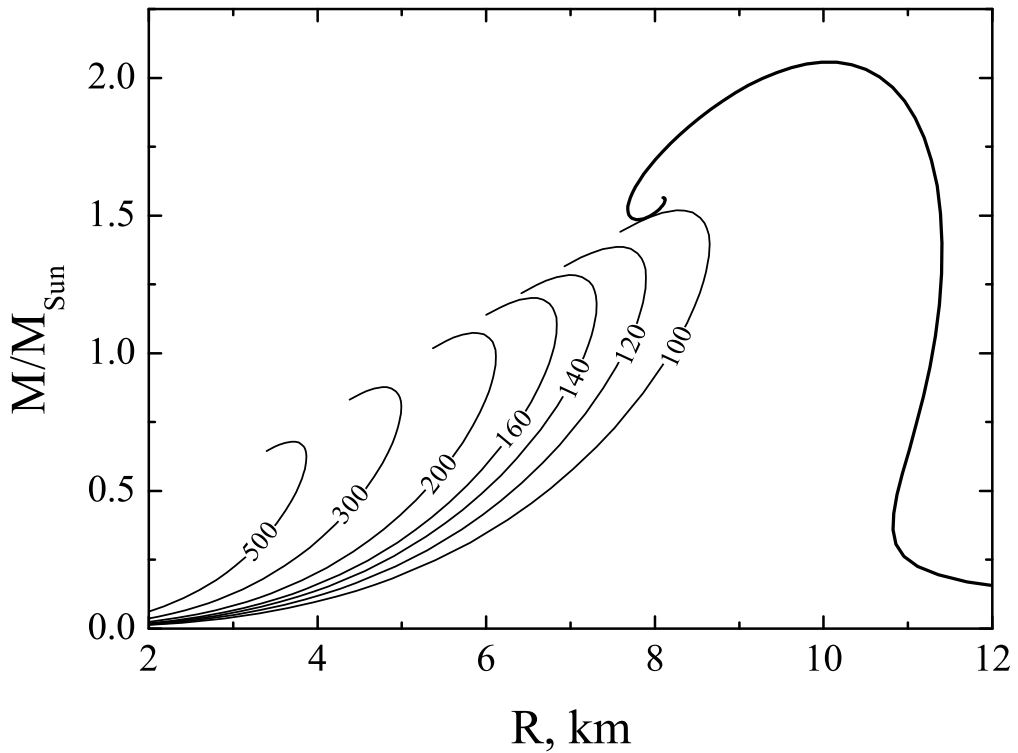} \caption{\rm
Mass–radius diagram for a star made of ordinary matter (thick
line) and purely quark stars (thin lines). The numbers at the
lines indicate the parameter $B$.} \label{Pix-PureQuark}
\end{center}
\end{figure}
The diagrams relating the stellar mass $M$ and radius $R$ are
applied to study the parameters of stellar models, in particular,
their stability. An example of such a diagram is shown in
Fig.~\ref{Pix-PureQuark}. The thick line indicates the mass–radius
relation for stars made of ordinary matter, without any phase
transition. The thin lines correspond to purely quark stars
containing no ordinary matter with their surface pressure
$\WID{P}{s}=0$ and density $\WID{\rho}{s}\neq 0$. The numbers at
the lines indicate the parameter $B$ in units of
$\mbox{MeV/fm}^3$. A characteristic feature of the diagrams for
quark stars is their passage through the coordinate origin $M=0,\
R=0$. It should also be noted that all these mass–radius curves
for quark stars are similar to one another (see the Section
“Dimensionless Form of the Equations” below).

\begin{figure}[htb]
\begin{center}
\epsfxsize=14cm \hspace{-0cm}\epsffile{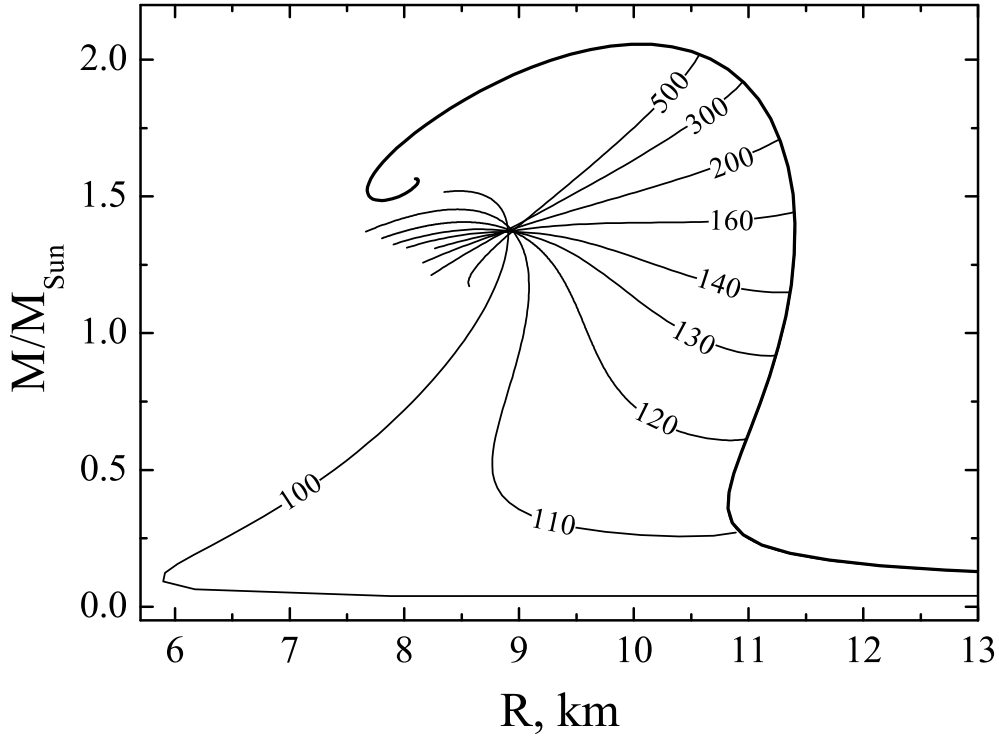}
\caption{\rm Mass--radius diagram of hybrid stars for various
values of the parameter $B$} \label{Pix-Magic_Point_main}
\end{center}
\end{figure}
The mass–radius diagram for hybrid (i.e., containing both phases)
stars calculated for our EOS is shown in
Fig.~\ref{Pix-Magic_Point_main}. The thick line again indicates
the dependence $M(R)$ for the EOS without any phase transition to
quark matter. The thin lines indicate these dependencies for
various values of the parameter $B$ (the values of $B$ are
indicated by the numbers in units of $\mbox{MeV/fm}^3$). The
density at which the phase transition begins $\rho_1$ is uniquely
related to $B$. This dependence is approximately described by the
formula (see Yudin et al. 2013)
$\rho_1/\WID{\rho}{n}=-3+1.5\ln(B{-}91)$, where
$\WID{\rho}{n}\approx 2.6{\times}10^{14}~\mbox{g/cm}^3$ --- is the
nuclear density and $B$ is measured in units of $\mbox{MeV/fm}^3$.
For example, $B=120$ corresponds to $\rho_1\approx
2\WID{\rho}{n}$, while for $B=145$ we have $\rho_1\approx
3\WID{\rho}{n}$. The curve with $B=100$ at $M\gtrsim\NMS{0.1}$
describes an almost pure quark star with a thin crust made of
ordinary matter and, therefore, exhibits a dependence $M(R)$
typical of such stars. On the other hand, as can be seen from the
figure, all stars with quark cores at $B\gtrsim 160$ are unstable.
Naturally, these specific values are unique to our model EOS.

Let us now turn to the formulation of the problem. As can be seen
from Fig.~\ref{Pix-Magic_Point_main}, all curves with different B
intersect in a very narrow region on the $(M{-}R)$ diagram (but
not at a point!). This property, which is surprising per se, not
only leads to some interesting consequences that we will discuss
in conclusion but also undoubtedly requires an explanation.
Actually, our paper is devoted to this explanation. Note also that
such a behavior of the curves $M(R)$ is not a unique property of
precisely our EOS. the same effect can be seen, for example, in
Fig.~15 from Schertler et al. (2000), in Fig.~4 from Fraga et al.
(2002) and in Fig.~4 from Sagert et al. (2009).

Before turning to the main part of our work, we will emphasize
once again the model status of our EOS. At present, the existence
of neutron stars with a mass $M\approx 2M_\odot$ has been firmly
established from observations (Demorest et al. 2010). As can be
seen from Fig.~\ref{Pix-Magic_Point_main}, our EOS gives $M\approx
1.5 M_\odot$ for the maximum mass of hybrid stars. Constructing
the models of hybrid stars that satisfy observations is a
separate, complex but accomplishable task (see, e.g., Weissenborn
et al. 2011). For our purposes, it will suffice that the EOS used
convey correctly the main characteristic properties of hybrid
stars. In particular, we will show below that the linearity of the
EOS for quark matter that is postulated in the bag model
(Eq.~(\ref{P-E-B})) but is also valid with a good accuracy in more
sophisticated models (see, e.g., Zdunik and Haensel 2013; Bombaci
and Logoteta 2013), which is needed for the existence of a
“special point”, turns out to be a decisive property. In our view,
the old result by Rhoades and Ruffini (1974), who found through
variational calculations that precisely the linear EOS of the core
maximizes the maximum neutron star mass for a known EOS of the
crust, is remarkable in this context.

\section*{DERIVATION OF THE MAIN CONDITION}
\begin{figure}[htb]
\begin{center}
\epsfxsize=14cm \hspace{-0cm}\epsffile{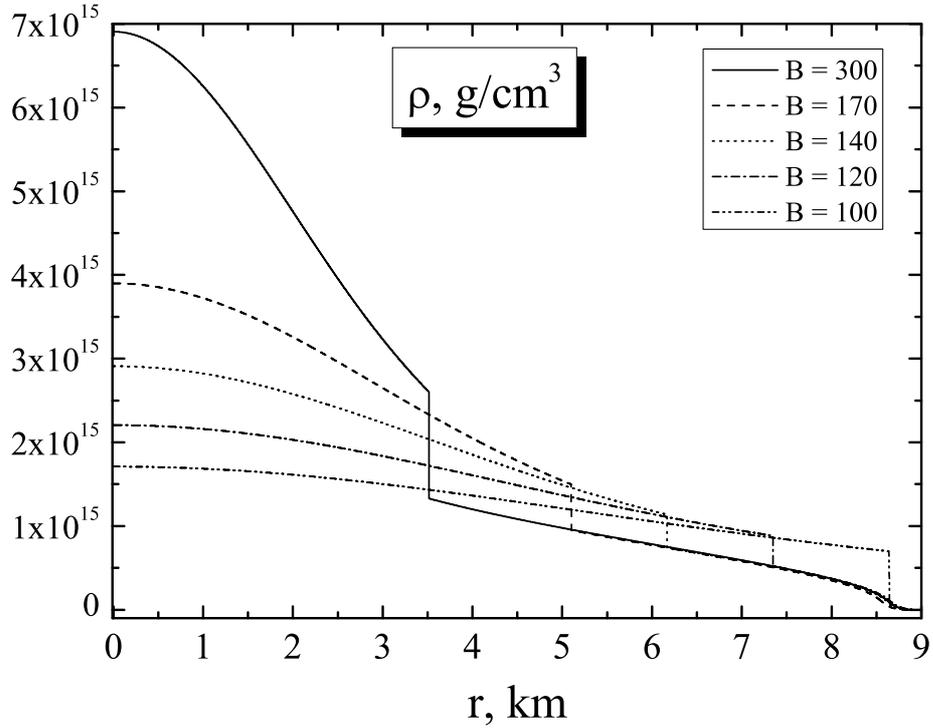} \caption{\rm
Density $\rho$ versus radial coordinate $r$ in a star near the
point of intersection for several values of the parameter $B$}
\label{Pix-Rho_r}
\end{center}
\end{figure}

 To come close to understanding the causes of the above
effect, we need to compare the structures of stars near the point
of intersection in Fig.~\ref{Pix-Magic_Point_main}. These stars
corresponding to different values of the parameter B should have
similar masses and radii. In Fig.~\ref{Pix-Rho_r} the baryon
density of matter $\WID{\rho}{b}$
($\WID{\rho}{b}\equiv\nucmu\WID{n}{b}$, where $\WID{n}{b}$ is the
baryonic charge density and $\nucmu$ is the atomic mass unit) is
plotted against the radial coordinate $r$. For each given value of
$B$, we chose a star near the point of intersection. As can be
seen, these stars have a virtually identical crust made of
ordinary matter to which a quark core is “stitched” at different
depths, depending on the parameter $B$. For example, the
transition occurs at $r\approx 3.5$~km for $B = 300$, at $r\approx
5$~km for $B = 170$, etc. Thus, when changing the parameter $B$,
the quark matter–-ordinary matter boundary is shifted, leaving the
crust virtually unchanged. Let us formalize this condition.

Let us first write the stellar equilibrium equations under general
relativity conditions (the Tolman–Oppenheimer–Volkoff equations):
\begin{align}
\frac{d P}{d r}&=-\frac{G(P+\epsilon)(m+\frac{4\pi r^3}{c^2}P)}{c^2 r(r-\frac{2 G m}{c^2})},\label{dPdr}\\
\frac{d m}{d r}&=\frac{4\pi r^2}{c^2}\epsilon,\label{dmdr}
\end{align}
where $m$ is the total (gravitating) mass within a sphere of
radius $r$. We will now denote the quantities referring to
ordinary and quark matter by the subscripts 1 and 2, respectively.
The phase equilibrium conditions at the boundary are reduced to
the equality of the matter pressures and chemical potentials
(everywhere below, we set the temperature equal to zero):
\begin{gather}
P_1(n_1)=P_2(n_2),\label{PhaseEquilibrium_P}\\
\WID{\mu}{pt}=\frac{P_1+\epsilon_1}{n_1}=\frac{P_2+\epsilon_2}{n_2}.\label{PhaseEquilibrium_mu}
\end{gather}
Recall that $n$ is the baryonic charge density. The EOS in the
second phase can be written as
\begin{gather}
\epsilon_2=\epsilon_2(\WID{n}{2},\zeta),\\
P_2=\WID{n}{2}^2\frac{\partial}{\partial\WID{n}{2}}\!\!\left(\frac{\epsilon_2}{\WID{n}{2}}\right),
\end{gather}
where $\zeta$ is some parameter (in the case of quark matter, it
is uniquely related to $B$). A change in $\zeta$ leads to a change
in the phase equilibrium parameters $\delta n_1$ and $\delta n_2$,
which are determined by varying Eqs. (\ref{PhaseEquilibrium_P})
and (\ref{PhaseEquilibrium_mu}):
\begin{align}
\triangle P&=\DER{P_1}{n_1}\delta n_1=\DER{P_2}{n_2}\delta
n_2+n_2^2\frac{\partial}{\partial
n_2}\!\!\left(\frac{1}{n_2}\DER{\epsilon_2}{\zeta}\right)\delta\zeta,\label{PE_P_var}\\
\triangle\WID{\mu}{pt}&=\DER{P_1}{n_1}\frac{\delta
n_1}{n_1}=\DER{P_2}{n_2}\frac{\delta
n_2}{n_2}+\frac{\partial^2\epsilon_2}{\partial
n_2\partial\zeta}\delta\zeta.\label{PE_mu_var}
\end{align}
Eliminating $\delta n_1$ and $\delta n_2$ from Eqs.
(\ref{PE_P_var}) and (\ref{PE_mu_var}) , we will find the relation
between the change in pressure at the phase equilibrium point
$\triangle P$ and the change in $\zeta$:
\begin{equation}
\triangle
P=\frac{\delta\zeta}{\lambda{-}1}\DERS{\epsilon_2}{\zeta}{},\label{dPzeta}
\end{equation}
where we denote $\lambda\equiv n_2/n_1$.

Let us now return to our star and suppose that it lies in the
region where the curves in Fig.~\ref{Pix-Magic_Point_main}
intersect. A change in $\zeta$ in phase 2 causes the phase
boundary at $r=r_0$ to be shifted by $\delta r$; in this case,
according to the condition $M,R=\mathrm{const}$, only the central
region with phase 2 changes, while the crust at $r>r_0+\delta r$
remains unchanged. The change in pressure at the phase boundary
can then be found as
\begin{equation}
\triangle P=\left(\frac{d P}{d r}\right)_{\!\!1}\delta
r=\frac{1}{\lambda}\left(\frac{d P}{d r}\right)_{\!\!2}\delta
r,\label{gradP}
\end{equation}
where the pressure gradients are found from the
Tolman–Oppenheimer–Volkoff equation (\ref{dPdr}). The last
equality in (\ref{gradP}) follows from (\ref{dPdr}) and the phase
equilibrium condition (\ref{PhaseEquilibrium_mu}). Similarly, the
change in the mass coordinate $m_0$ of the phase boundary in the
star is
\begin{equation}
\triangle m=\frac{4\pi r_0^2}{c^2}\epsilon_1\delta r=\frac{4\pi
r_0^2}{c^2}\left[\frac{P_2+\epsilon_2}{\lambda}-P_2\right]\delta
r=\left(\frac{d m}{d
r}\right)_{\!\!2}\left[1-(\lambda{-}1)\frac{P_2}{\epsilon_2}\right]\frac{\delta
r}{\lambda},\label{gradm}
\end{equation}
where we again used Eqs. (\ref{PhaseEquilibrium_P}) and
(\ref{PhaseEquilibrium_mu}). The changes in the pressure and mass
coordinate of the boundary of the core with phase 2 can also be
found as
\begin{gather}
\triangle P=\left(\frac{d P}{dr}\right)_{\!\!2}\delta
r+\DERS{P}{\WID{P}{c}}{r,\zeta}\!\!\delta
\WID{P}{c}+\DERS{P}{\zeta}{r,\WID{P}{c}}\!\!\delta\zeta,\label{dPseries}\\
\triangle m=\left(\frac{d m}{dr}\right)_{\!\!2}\delta
r+\DERS{m}{\WID{P}{c}}{r,\zeta}\!\!\delta
\WID{P}{c}+\DERS{m}{\zeta}{r,\WID{P}{c}}\!\!\delta\zeta.\label{dmseries}
\end{gather}
Here, the first term is attributable to the change in core radius,
the second term is attributable to the change in central pressure
$\WID{P}{c}$ and to the coordinated change in pressure at all
points of the core caused by it, and the last term is attributable
to the change in $\zeta$ in the EOS of the central phase. We can
now bring together the equations for $\triangle P$ (\ref{dPzeta}),
(\ref{gradP}), (\ref{dPseries}) and $\triangle m$ (\ref{gradm}),
(\ref{dmseries}) and obtain a system of three equations for
$\delta r$, $\delta\zeta$ and $\delta\WID{P}{c}$:
\begin{align}
\frac{dP}{dr}\left[\frac{\lambda{-}1}{\lambda}\right]\delta r &=
\DERS{\epsilon}{\zeta}{}\delta\zeta,\label{Ur1}\\
-\frac{dP}{dr}\left[\frac{\lambda{-}1}{\lambda}\right]\delta r &=
\DERS{P}{\WID{P}{c}}{r,\zeta}\!\!\delta
\WID{P}{c}+\DERS{P}{\zeta}{r,\WID{P}{c}}\!\!\delta\zeta,\label{Ur2}\\
-\frac{dm}{dr}\left[\frac{P+\epsilon}{\epsilon}\right]\left[\frac{\lambda{-}1}{\lambda}\right]\delta
r &=\DERS{m}{\WID{P}{c}}{r,\zeta}\!\!\delta
\WID{P}{c}+\DERS{m}{\zeta}{r,\WID{P}{c}}\!\!\delta\zeta.\label{Ur3}
\end{align}
Since all of the quantities considered, except the parameter
$\lambda$, refer to the second (central) phase, we omitted the
subscript 2 here for brevity. For these equations to have a
nonzero solution, the determinant of the system must become zero.
This condition gives us the main equation
\begin{equation}
\DERS{P}{\WID{P}{c}}{r,\zeta}\left[\frac{dm}{dr}
\frac{P+\epsilon}{\epsilon\frac{dP}{dr}}\DERS{\epsilon}{\zeta}{}+\DERS{m}{\zeta}{r,\WID{P}{c}}\right]=
\DERS{m}{\WID{P}{c}}{r,\zeta}\left[\DERS{\epsilon}{\zeta}{}+\DERS{P}{\zeta}{r,\WID{P}{c}}\right].\label{MainEquation}
\end{equation}
All of the quantities in this equation refer to the central phase
(phase 2), because the parameter $\lambda$ relating the phases
dropped out of it. This remarkable fact implies that the property
to conserve the total stellar mass and radius as the core size
changes is determined only by the central phase and does not
depend directly on the crust parameters! If condition
(\ref{MainEquation}) is met at some point of the star and if this
point is the phase transition point (i.e., Eqs.
(\ref{PhaseEquilibrium_P}) and (\ref{PhaseEquilibrium_mu}) hold at
it), then the total stellar mass and radius will not change at
small variations in the parameter $\zeta$ of the central phase.

\section*{DIMENSIONLESS FORM
OF THE EQUATIONS}
 Let us now turn again to the case of
stars with quark cores. As we have seen, the EOS for quark matter
in the simplest case is a special case of the linear EOSs:
$P=\alpha(\epsilon{-}\epsilon_0)$ with $\alpha=1/3$ and
$\epsilon_0=4B$. This fact allows the Tolman–Oppenheimer–Volkoff
equilibrium equations (\ref{dPdr}) and (\ref{dmdr}) to be made
dimensionless (for more details, see Haensel et al. 2007). More
specifically, let us introduce dimensionless variables
$\rho\equiv\epsilon/\epsilon_0$, $x=r/\WID{r}{dim}$ and
$\mu=m/\WID{m}{dim}$, with $\WID{r}{dim}=c^2/\sqrt{4\pi
G\epsilon_0}$ and $\WID{m}{dim}=c^4/G\sqrt{4\pi G\epsilon_0}$; in
this case, $P=\alpha\epsilon_0(\rho{-}1)$. The equilibrium
equations (\ref{dPdr}) and (\ref{dmdr}) will then be written as
\begin{align}
\alpha\frac{d\rho}{dx}&=-\left[\rho+\alpha(\rho{-}1)\right]\frac{\mu+x^3\alpha(\rho{-}1)}{x(x{-}2\mu)},\label{dPdr_dim}\\
\frac{d\mu}{dx}&=x^2\rho.\label{dmdx_dim}
\end{align}
Having specified some central value of $\rho(0)=\WID{\rho}{c}\geq
1$, $\mu(0)=0$, we can integrate these equations to the point
$\rho=1$, representing the surface of a quark star ($P=0$). At
fixed $\alpha$ we obtain a family of solutions with the parameter
$\WID{\rho}{c}$.

Let us now rewrite the main equation (\ref{MainEquation}) in
dimensionless variables. Suppose that $\zeta=\epsilon_0$. Given
that $B=\epsilon_0\alpha/(1{+}\alpha)$, we will then obtain
\begin{equation}
\DERS{\epsilon}{\zeta}{}=\frac{\alpha}{1{+}\alpha}.
\end{equation}
The derivatives with respect to the central pressure are
\begin{align}
\DERS{P}{\WID{P}{c}}{r,\epsilon_0} &= \DERS{\rho}{\WID{\rho}{c}}{x},\\
\DERS{m}{\WID{P}{c}}{r,\epsilon_0} &=
\frac{\WID{m}{dim}}{\alpha\epsilon_0}\DERS{\mu}{\WID{\rho}{c}}{x}.
\end{align}
Finally, the derivatives with respect to $\zeta$ are expressed as
\begin{align}
\DERS{P}{\zeta}{r,\WID{P}{c}} \!\!&= \alpha\left[\rho{-}1+\frac{x}{2}\frac{d\rho}{dx}-\DERS{\rho}{\WID{\rho}{c}}{x}(\WID{\rho}{c}{-}1)\right],\\
\DERS{m}{\zeta}{r,\WID{P}{c}} \!\!&=
\frac{\WID{m}{dim}}{\epsilon_0}\left[-\frac{\mu}{2}+\frac{x}{2}\frac{d\mu}{dx}-\DERS{\mu}{\WID{\rho}{c}}{x}(\WID{\rho}{c}{-}1)\right].
\end{align}
Gathering all these expressions and replacing $\frac{d\mu}{dx}$ by
its value from (\ref{dmdx_dim}), we will obtain our main equation
(\ref{MainEquation}) in dimensionless form:
\begin{equation}
\DERS{\mu}{\WID{\rho}{c}}{x}\!\!\left[\rho-
\frac{\alpha}{1{+}\alpha}+\frac{x}{2}\frac{d\rho}{dx}\right]\frac{d\rho}{dx}=
\DERS{\rho
}{\WID{\rho}{c}}{x}\!x^2\!\left[\rho-\frac{\alpha}{1{+}\alpha}+\frac{x}{2}\frac{d\rho}{dx}\left(\rho-\frac{\mu}{x^3}\right)\right],\label{MainEquation_dim}
\end{equation}
where $\frac{d\rho}{dx}$ can be determined from
Eq.~(\ref{dPdr_dim}).

\pagebreak
\section*{HOMOLOGOUS VARIABLES}
To analyze Eq. (\ref{MainEquation_dim}) we will have to make a
small digression. It is well known from the theory of polytropes
that the system of stellar equilibrium equations (\ref{dPdr}) and
(\ref{dmdr}) in the Newtonian limit with a polytropic EOS can be
transformed to $x\frac{dv}{dx}=\WID{f}{v}(u,v)$ and
$x\frac{du}{dx}=\WID{f}{u}(u,v)$ by introducing the so-called
homologous variables $(u, v)$. These equations are reduced to one
differential equation
$\frac{dv}{du}=f(u,v)=\frac{\WID{f}{v}(u,v)}{\WID{f}{u}(u,v)}$
(see Chandrasekhar 1950). In this case, all solutions of the
system with different central pressures (densities) fall on the
same curve in the $(u, v)$ plane. It turns out that for an EOS of
the form $P=\alpha\epsilon$, the stellar equilibrium equations can
also be similarly transformed within the framework of general
relativity by introducing Milne’s homologous variables (see
Chandrasekhar 1972; Chavanis 2002). However, the additional term
$\epsilon_0$ in our expression $P=\alpha(\epsilon{-}\epsilon_0)$
violates homology. Nevertheless, we managed to find the variables
in which the equilibrium equations (\ref{dPdr_dim}) and
(\ref{dmdx_dim}) with the EOS $P=\alpha\epsilon_0(\rho{-}1)$ are
approximately homologous, i.e., their solutions fall
\emph{virtually} on the same curve in some domain of variables
$(u,v)$ for moderately large $\alpha$ (recall that $\alpha=1/3$ in
our case). Thus, let us introduce the variables $u$ and $v$:
\begin{align}
v &= -\frac{\alpha x}{\rho+\alpha(\rho{-}1)}\frac{d\rho}{dx}
=\frac{\mu+x^3\alpha(\rho{-}1)}{x{-}2\mu},\label{v_definition}\\
u &= \frac{x^3\left[\rho+\alpha(\rho{-}1)\right]}{3\mu+\alpha
x^3(\rho{-}1)}.\label{u_definition}
\end{align}
The central point of the star corresponds to $v=0, u=1$. The
equations for $v$ and $u$ are:
\begin{align}
x\frac{dv}{dx} &= \frac{1+2v}{1{+}2 x^2\alpha(\rho{-}1)}\left[x^2(\rho+3\alpha(\rho{-}1))-v+vx^2(\rho-\alpha(\rho{-}1))\right],\label{dvdx_or}\\
\frac{x}{u}\frac{du}{dx} &=
3-\frac{1+\alpha}{\alpha}v-u(3-v).\label{dudx}
\end{align}
Naturally, only the second equation has the necessary homologous
form. However, the first equation can also be brought to a
homologous form in the limiting cases. First, let $\rho\gg 1$.
This corresponds to $\epsilon\gg\epsilon_0$, i.e.
$P\approx\alpha\epsilon$, the case where, according to what has
been said above, a homologous solution definitely exists. To
within terms $o\!\left(\frac{1}{\rho}\right)$ instead of Eq.
(\ref{dvdx_or}) we then have
\begin{equation}
\frac{x}{v}\frac{dv}{dx} = 3u(1{+}v)-(1{+}2v)+6\alpha
u\frac{1-u(1{+}v)}{1+\alpha(1{+}2u)},\label{dvdv_high}
\end{equation}
where we expressed $\mu$ and $x$ in terms of $u$, $v$ and $\rho$
using definitions (\ref{v_definition}) and (\ref{u_definition}).
The third term in this expression containing the factor $\alpha$,
is definitely small at the beginning of the homologous curve at
$u\approx 1$ and $v\approx 0$, where $u\approx 1{-}v/5\alpha$.

\noindent Consider the other limiting case of $\rho\approx 1$. To
within $o(\rho{-}1)$, we then have
\begin{equation}
\frac{x}{v}\frac{dv}{dx} = 3u(1{+}v)-(1{+}2v).\label{dvdv_low}
\end{equation}
As we see, this expression coincides with the first two terms in
(\ref{dvdv_high}). It is also interesting to note that the next
expansion term, of order $O(\rho{-}1)$, is $6\alpha
u(\rho{-}1)[1-u(1{+}v)]$.

\begin{figure}[htb]
\begin{center}
\epsfxsize=14cm \hspace{-0cm}\epsffile{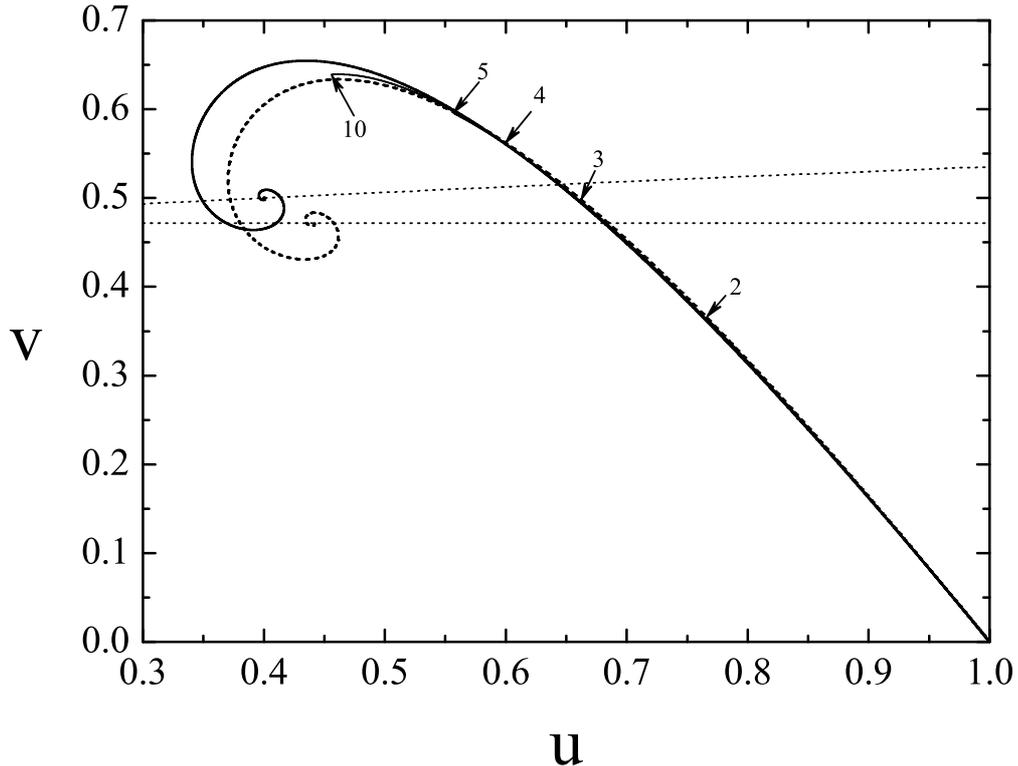} \caption{\rm
Stellar structure in homologous variables. The solid thick and
dashed lines represent the limits $\rho\gg 1$ and $\rho\sim 1$
respectively.} \label{Pix-UV}
\end{center}
\end{figure}
\noindent The thick solid spiral line in Fig.~\ref{Pix-UV}
indicates the result of our calculation according to
Eqs.~(\ref{dudx}) and (\ref{dvdv_high}) (the limit $\rho\gg 1$)
and the dashed spiral (the limit $\rho\sim 1$) corresponds to the
solution according to (\ref{dudx}) and (\ref{dvdv_low}). The
structure of real quark stars (corresponding to the solution of
Eqs.~(\ref{dPdr_dim}) and (\ref{dmdx_dim})) in homologous
variables is indicated by the thin solid lines almost coincident
with the spiral ones. The arrows indicate the points corresponding
to the surface ($\rho=1$) of these stars; the number at the arrow
indicates the corresponding dimensionless central density
$\WID{\rho}{c}$. As can be seen, all stars have a similar
homologous structure in much of the $(u,v)$ plane; deviations are
observed only in the region of the spiral turn. In this sense, our
variables $(u,v)$ are actually “almost homologous”. The meaning of
the thin dotted lines will be discussed below.

\section*{SOLUTION OF THE MAIN EQUATION}
Let us now return to our main equation (\ref{MainEquation_dim}),
which expresses the condition for the total stellar mass and
radius being constant at small variations in the parameter of the
central phase in dimensionless variables. The main problem is to
find the derivatives
$\left(\frac{\partial\mu}{\partial\WID{\rho}{c}}\right)_x$ and
$\left(\frac{\partial\rho}{\partial\WID{\rho}{c}}\right)_x$. Let
us relate the quantities $\mu$ and $\rho$ to the homologous
variables $u$ and $v$:
\begin{align}
\frac{\mu}{x}&=\frac{v(1+\alpha(1{-}u))+\alpha
x^2}{(1{+}\alpha)(1{+}2v)+2\alpha u(1{-}v)},\label{mu_u_v_rho}\\
x^2(\rho{-}1)&=\frac{3uv-(1{+}2v)x^2}{(1{+}\alpha)(1{+}2v)+2\alpha
u(1{-}v)}.\label{x_u_v_rho}
\end{align}
A change in the central density $\WID{\rho}{c}$ at
$x=\mbox{const}$ leads to a change in the parameters $u$ and $v$.
However, no matter what this change is, it is just reduced to some
shift along the homologous curve defined by the solution of the
equation $dv/du=f(u,v)$. Thus, we can write
$\delta\mu=\left[\DERS{\mu}{u}{}+\DERS{\mu}{v}{}\frac{\WID{f}{v}}{\WID{f}{u}}\right]\delta
u$ and
$\delta\rho=\left[\DERS{\rho}{u}{}+\DERS{\rho}{v}{}\frac{\WID{f}{v}}{\WID{f}{u}}\right]\delta
u$, where the functions $\WID{f}{v}$ and $\WID{f}{u}$ are
determined from Eqs.~(\ref{dvdx_or}) and (\ref{dudx}).
Substituting this into the main equation (\ref{MainEquation_dim}),
we obtain a cumbersome expression that, however, is simplified
after some transformations to
\begin{equation}
u_{*}=\left[v_{*}^{2}(3{+}\alpha)+v_{*}(3{-}\alpha)-6\alpha\right]
\frac{1+(1{+}\alpha)(\rho_{*}{-}1)}{4\alpha^2(\rho_{*}{-}1)(1{-}v_{*})(3{-}v_{*})}.\label{u_star}
\end{equation}
Here and below, the asterisk marks the values of the quantities at
the “special point”. This relation specifies the sought-for
condition that the homologous variables $u_*$ and $v_*$ as well as
the parameter $\rho_*$ should satisfy to serve as the solution of
(\ref{MainEquation_dim}) (here, we expressed $x$ in the formulas
in terms of $u$, $v$, and $\rho$ using (\ref{x_u_v_rho})). In this
case, $u$ and $v$ should lie on the homologous curve. The
parameter $\rho$ on the $(u,v)$ diagram is a "hidden" variable,
i.e., different values of $\rho$ correspond to the same values of
$u$ and $v$.

Consider the limiting cases of Eq.~(\ref{u_star}). First, let
$\rho_*\rightarrow 1$, i.e., the phase transition occurs in the
crust, the star is virtually a purely quark one. The following
condition should then be met:
\begin{equation}
v_{*}^{2}(3{+}\alpha)+v_{*}(3{-}\alpha)-6\alpha=0,\label{dot_horizon}
\end{equation}
which for $\alpha=1/3$ leads to $v_*\approx 0.4718$. This limit is
indicated by the horizontal dotted line in Fig.~\ref{Pix-UV}. Its
intersection with the homologous curve gives the corresponding
$u_*\approx 0.685$. The other limiting case of $\rho_*\rightarrow
\infty$ gives an equation of the curve indicated by the oblique
dotted line in Fig.~\ref{Pix-UV}:
\begin{equation}
u=(1{+}\alpha)\frac{v^{2}(3{+}\alpha)+v(3{-}\alpha)-6\alpha}{4\alpha^2(1{-}v)(3{-}v)}.\label{dot_naklon}
\end{equation}
Its intersection with the homologous curve occurs at $v_*\approx
0.515$ and $u_*\approx 0.645$. Interestingly, these curves also
pass through the limiting points of the corresponding homologous
curves (the centers of the spirals corresponding to the solutions
of the equations $\WID{f}{u}(u,v)=0$ and $\WID{f}{v}(u,v)=0$ (see
(\ref{dudx}), (\ref{dvdv_low}) and (\ref{dvdv_high}))). For
example, the horizontal curve defined by Eq.~(\ref{dot_horizon})
also passes through the limiting point of the homologous curve for
$\rho\sim 1$ with $v\approx 0.4718$ and $u\approx 0.44$, while the
curve defined by Eq.~(\ref{dot_naklon}) passes through the
limiting point of the curve for $\rho\gg 1$ with coordinates
$v=2\alpha/(1{+}\alpha)=0.5$ and $u=(1{+}\alpha)/(3{+}\alpha)=0.4$
(the numerical values are indicated for $\alpha=1/3$).

Thus, all the states of interest to us lie in a small segment of
the homologous curve: from $(u\approx 0.685, v\approx 0.4718)$ to
$(u\approx 0.645, v\approx 0.515)$ (see Fig.~\ref{Pix-UV}). Each
point of this segment of the curve corresponds to some density
$\rho_*$ ((according to Eq.~(\ref{u_star})) between $\rho_*=1$ for
the first above pair $(u,v)$ and $\rho_*=\infty$ for the second
one. Accordingly, for each such point there exists such a unique
value of $\WID{\rho}{c}$ that having begun the integration of the
equilibrium equations (\ref{dPdr_dim}) and (\ref{dmdx_dim}) with
this central density, we end up at the point $(u_*,v_*)$ with the
required density $\rho_*$. If the phase diagram of matter is
structured in such a way that the phase transition occurs at this
point, then such a star will have the sought-for property: its
total mass and radius will not depend on small variations in the
parameter $B$ (or $\epsilon_0$) of the central phase.

\section*{LARGE SCALE}
\begin{figure}[htb]
\begin{center}
\epsfxsize=14cm \hspace{-0cm}\epsffile{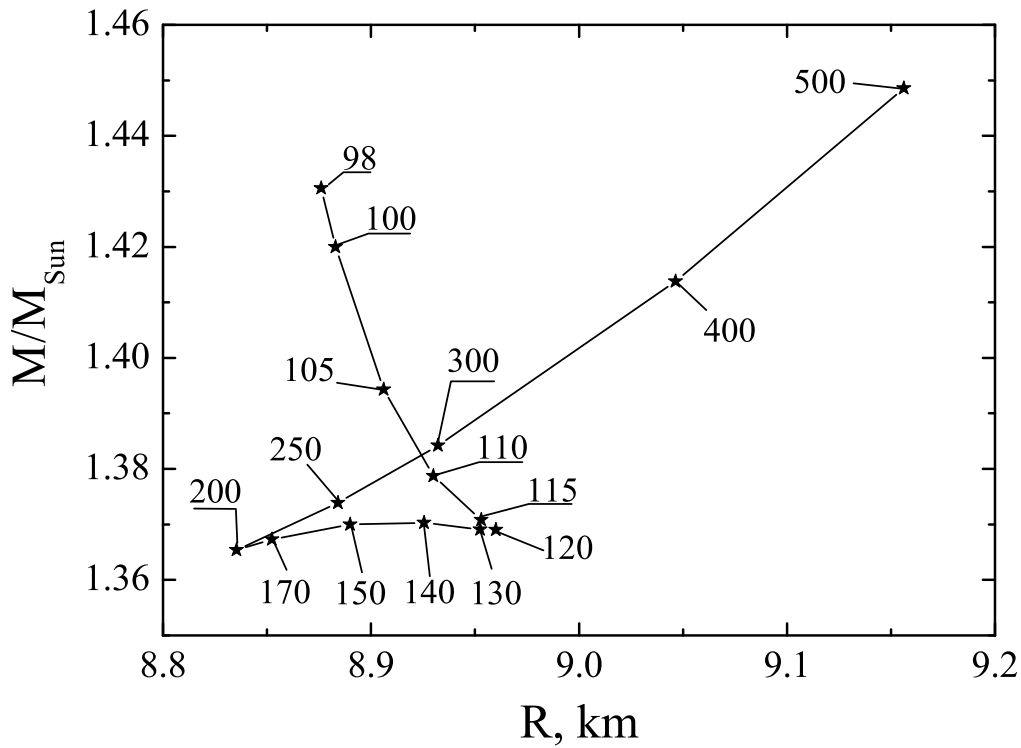} \caption{\rm
The line of stationary points. The numbers indicate the values of
B in units of MeV/fm$^3$} \label{Pix-MP_zoom}
\end{center}
\end{figure}

 The condition for the total stellar mass and radius
being constant (\ref{MainEquation}), its dimensionless form
(\ref{MainEquation_dim}) and corollary (\ref{u_star}) are local,
i.e., they are valid only at small variations in the parameter $B$
(or $\epsilon_0$) of the central phase. In this case, the curves
corresponding to various, slightly differing values of $B$, on the
mass– radius diagram intersect at a single point that we will call
a “stationary point”. Naturally, different coordinates of the
stationary points generally correspond to different values of $B$.
The line of stationary points on the mass–radius diagram is shown
in Fig.~\ref{Pix-MP_zoom}. The numbers denote the corresponding
values of $B$ in units of MeV/fm$^3$. As can be seen, this curve
has a rather peculiar shape. Owing to the two kinks at $B\approx
120$ and $B\approx 200$ the bulk of it occupies a bounded region
of the diagram. This is one of the reasons why the mass–radius
curves corresponding to different, even greatly differing values
of $B$, intersect in a small region (see
Fig.~\ref{Pix-Magic_Point_main}). The second reason is related to
the topology of the diagram: for example, the curves corresponding
to small $B$, whose stationary points lie above and to the left of
the central triangle of stationary points run from bottom to top
(as it should be for almost purely quark stars). The curves for
intermediate $B$ run from right to left, while those for large $B$
drop from top to bottom and, passing through their stationary
points, nevertheless also pass through the central zone of the
diagram. Let us try to understand the behavior of the line of
stationary points.

Consider a star with the parameters of the boundary of its quark
core satisfying condition (\ref{u_star}). Let us denote this
condition by the relation $G(u_*,v_*,\rho_*)=0$ and call the
parameters that satisfy it the parameters of the stationary point.
At a small change in $\epsilon_0$ and a corresponding change in
the central density $\WID{\rho}{c}$ the total stellar mass and
radius will remain unchanged. The core boundary now corresponds to
new values of the dimensionless parameters, $u'=u_*+\delta u$,
$v'=v_*+\delta v$ and $\rho'=\rho_*+\delta\rho$. If these values
also satisfy the condition $G(u',v',\rho')=0$, then we can further
change $\epsilon_0$, conserving the total stellar mass and radius,
etc. However, it is obvious that this is generally not the case
and the new parameters $\{u',v',\rho'\}$ need not be the
parameters of the stationary point. Let us derive the condition
under which the new state is also a stationary point. We have two
relations: $\delta v=\WID{f}{v}/\WID{f}{u}\delta u$ and
$\DERS{G}{u}{}\delta u+\DERS{G}{v}{}\delta
v+\DERS{G}{\rho}{}\delta\rho=0$, whence
\begin{gather}
\delta
u=-\frac{\WID{f}{u}\DERS{G}{\rho}{}\delta\rho}{\WID{f}{u}\DERS{G}{u}{}+\WID{f}{v}\DERS{G}{v}{}},\label{non-mov-u}\\
\delta
v=-\frac{\WID{f}{v}\DERS{G}{\rho}{}\delta\rho}{\WID{f}{u}\DERS{G}{u}{}+\WID{f}{v}\DERS{G}{v}{}},\label{non-mov-v}
\end{gather}
where, as has already been said, the function
$G(u_*,v_*,\rho_*)=0$ is determined from Eq.~(\ref{u_star}).

Consider now how the parameters $\{u,v,\rho\}$ actually change
during a shift that leaves the total stellar mass and radius
unchanged. For this purpose, let us again return to
Eqs.~(\ref{dPzeta}), (\ref{gradP}) and (\ref{gradm}), which relate
the changes in pressure $\triangle P$, mass $\triangle m$ and
radius $\delta r$ to the variation in $\epsilon_0$. Writing them
in dimensionless form and eliminating
$\delta\zeta=\delta\epsilon_0$, we will obtain the relation
\begin{align}
&\Lambda_1\frac{\delta x^2}{2 x^2}=\left[\lambda+(1{+}\alpha)(\lambda{-}1)\frac{x}{2}\frac{d\rho}{dx}\right]\delta\rho,\\
&\Lambda_1\delta\left(\frac{\mu}{x}\right)=\left[x^2\left(1+(\rho{-}1)(1-\alpha(\lambda{-}1))\right)-\lambda\frac{\mu}{x}\right]\delta\rho,
\end{align}
where we introduced the factor
\begin{equation}
\Lambda_1\equiv
x\frac{d\rho}{dx}\left[1-(1{+}\alpha)(\lambda{-}1)(\rho{-}1)\right].
\end{equation}
The quantities $x\frac{d\rho}{dx}$, $x^2$ and $\mu/x$ are
expressed in terms of $u$, $v$ and $\rho$ using
Eqs.~(\ref{v_definition}), (\ref{mu_u_v_rho}) and
(\ref{x_u_v_rho}). To write the result in a compact form, let us
split the quantities $\WID{f}{u}$ and $\WID{f}{v}$ as
$\WID{f}{u}=\WID{f}{u1}+\WID{f}{u2}$ and
$\WID{f}{v}=\WID{f}{v1}+\WID{f}{v2}$, where (see also
Eqs.~(\ref{dvdx_or}) and (\ref{dudx}))
\begin{align}
\WID{f}{u1}&=-u^2(3-v),\\
\WID{f}{u2}&=u\left[3-\frac{1{+}\alpha}{\alpha}v\right],\\
\WID{f}{v1}&=uv(1{+}v)\left[3-\frac{6\alpha
u(\rho{-}1)}{1{+}(\rho{-}1)\left[1{+}\alpha(1{+}2u)\right]}\right],\\
\WID{f}{v2}&=-v\left[1{+}2v-\frac{6\alpha
u(\rho{-}1)}{1{+}(\rho{-}1)\left[1{+}\alpha(1{+}2u)\right]}\right].
\end{align}
Here, in Eq.~(\ref{dvdx_or}) we expressed $x$ in terms of $u$, $v$
and $\rho$ using (\ref{x_u_v_rho}). We can now ultimately write
the resulting relations in a compact form:
\begin{align}
\delta u&=-\left(\WID{f}{u1}+\lambda\WID{f}{u2}\right)\frac{\alpha
\delta\rho}{v\Lambda_2},\\
\delta v&=-\left(\WID{f}{v1}+\lambda\WID{f}{v2}\right)\frac{\alpha
\delta\rho}{v\Lambda_2},
\end{align}
where we introduced the common factor
$\Lambda_2=\left[1{+}(\rho{-}1)(1{+}\alpha)\right]\left[1{-}(\rho{-}1)(1{+}\alpha)(\lambda{-}1)\right]$.
These equations define how the dimensionless variables $u$, $v$
and $\rho$ change during a shift that leaves the total stellar
mass and radius unchanged. They should be compared with
Eqs.~(\ref{non-mov-u}) and (\ref{non-mov-v}), which define the
shift between two stationary points. The requirement $\delta
u/\delta v=\WID{f}{u}/\WID{f}{v}$ immediately leads us to Eq.
(\ref{u_star}). This means that if we are at a stationary point,
then the shift will always be along the homologous curve
irrespective of $\lambda$. The second equation leads us to a
condition for $\lambda$:
\begin{equation}
\lambda=\frac{u(3{-}v)\left[(1{+}v)^2(3{-}v){+}8(1{-}v^2)\alpha-(3{-}v)^3\alpha^2\right]}
{(1{+}v)(7v^2{-}6v{+}3){+}8(1{-}v^2)(3{-}v)\alpha-(3{-}v)^3\alpha^2}.\label{lambda_condition}
\end{equation}
If the jump in density $\lambda$ satisfies condition
(\ref{lambda_condition}), then the stationary point also remains
stationary after the shift, i.e., the condition for the total
stellar mass and radius being constant becomes global. Otherwise,
when passing from one stationary point to another, the total mass
and radius will slightly change. For our case, $\alpha=1/3$, and
the parameters $u$ and $v$ lie within a narrow range from
$(u\approx 0.685, v\approx 0.4718)$ to $(u\approx 0.645, v\approx
0.515)$.  $\lambda\approx 1.664$ and $\lambda\approx 1.635$
respectively, correspond to them.

\begin{figure}[htb]
\begin{center}
\epsfxsize=14cm \hspace{-0cm}\epsffile{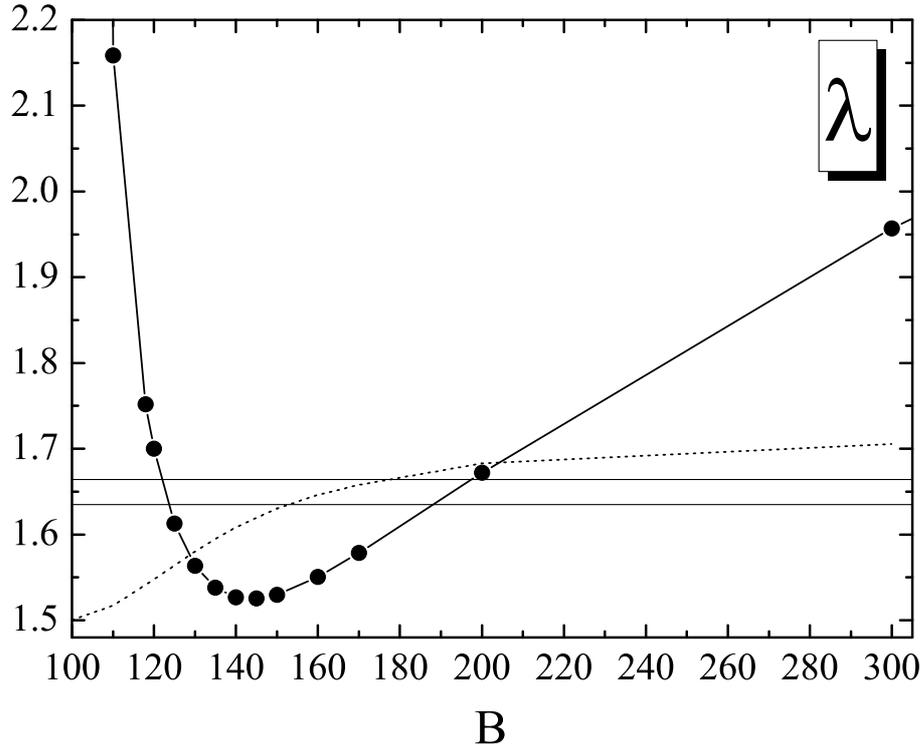} \caption{\rm
Jump in density $\lambda$ versus parameter $B$ for our EOS. The
solid horizontal lines indicate the special values according to
(\ref{lambda_condition}); the dotted line indicates the critical
value of $\lambda$ for the star to be stable.} \label{Pix-Lambda}
\end{center}
\end{figure}
How does the dependence of $\lambda$ on $B$ look in our case?
Figure~\ref{Pix-Lambda} gives the answer. It also shows the
special values of $\lambda$ listed above (solid horizontal lines)
and the critical value of $\lambda$ in the sense of the star’s
stability (dotted line) that we will briefly discuss below. As can
be seen, the narrow range of special values of $\lambda$ breaks up
the plot into several regions: in the zone $B\lesssim 122$,
$\lambda$ is larger than the special value and the line of
stationary points runs downward on the mass– radius diagram (see
Fig.~\ref{Pix-MP_zoom}). The region $B\approx 122\div 124$ is
special; here the property of stationarity is global, while the
mass and radius are almost constant. This zone is the turning
point in Fig.~\ref{Pix-Lambda}: further out, up to $B\approx 189$,
$\lambda$ is smaller than the special value and the line of
stationary points in Fig.~\ref{Pix-MP_zoom} changes its direction:
now the radius drops, while the mass changes little with
increasing $B$. The region $B\approx 189\div 199$ is again special
and the turning one in Fig.~\ref{Pix-MP_zoom}. As $B$ increases
further, $\lambda$ increases and the line of stationary points
runs monotonically to the upper right on the mass--radius diagram.

In conclusion, it remains for us to investigate two questions.
First, what determines the coordinates of the “special point”,
i.e., the characteristic mass $M_*$ and radius $R_*$ of the
intersection region? As a reference point, we will take an almost
purely quark star with $B_*\approx 100~\mbox{MeV/fm}^3$
($\epsilon_0=4B$); the mass and thickness of the crust made of
ordinary matter may be neglected (see Fig.~(\ref{Pix-Rho_r})).
Since $\rho\approx 1$, at its boundary, we can write
Eqs.~(\ref{v_definition}) and ($\ref{u_definition}$) in the
following form by substituting the numerical values:
\begin{align}
v_*&\approx 0.4718=\frac{\mu_*}{x_*{-}2\mu_*},\\
u_*&\approx 0.685=\frac{x_{*}^{3}}{3\mu_*}.
\end{align}
Hence, passing to dimensional units, we will obtain the following
characteristic values:
\begin{align}
R_* &=\WID{r}{dim}x_*=\frac{c^2 x_*}{\sqrt{4\pi
G\epsilon_0}}\approx
8.66~\mbox{km},\\
M_* &=\WID{m}{dim}\mu_*=\frac{c^4\mu_*}{G\sqrt{4\pi
G\epsilon_0}}\approx 1.42 M_\odot.
\end{align}
The total stellar radius will be slightly larger, because there is
also a tenuous “atmosphere” made of ordinary matter that makes
virtually no contribution to the total mass (see
Fig.~\ref{Pix-Rho_r}). The plot of stationary points
(Fig.~\ref{Pix-MP_zoom}) gives $\{1.38,8.9\}$ for the averaged
coordinates of the point of intersection
$\{M_*[M_\odot],R_*[\mbox{km}]\}$. It is interesting to compare
this quantity with the results obtained in other works: for
example, the point of intersection between themass–radius curves
in Fig. 15 from Schertler et al. (2000) gives $\{1.36,10\}$, Fig.
4 from Fraga et al. (2002) leads to $\{1.1,8\}$, and Fig. 4 from
Sagert et al. (2009) corresponds to $\{1.37,9.7\}$. If, however,
the approximation proposed in the book by Haensel et al. (2007) is
used for the EOS of the crust within the framework of our
approach, then we will obtain $\{1.37,9.12\}$ for the coordinates
of the special point.

The second question concerns the phase transition parameter
$\lambda\equiv n_2/n_1$. This parameter defines the stability of a
star when a new phase appears at its center: as Lighthill (1950)
showed, its critical value in the Newtonian limit is
$\WID{\lambda}{cr}=3/2$; at larger $\lambda$, stars with the phase
transition at their centers are hydrodynamically unstable. This
criterion was generalized to the case of general relativity by
Seidov (1971) and took the form
$\WID{\widehat{\lambda}}{cr}=3/2(1+P_*/\epsilon_1)$, where
$\widehat{\lambda}\equiv\epsilon_2/\epsilon_1$. It is easy to
derive the relation
\begin{equation}
\lambda=\widehat{\lambda}\frac{1+P_*/\epsilon_2}{1+\widehat{\lambda}P_*/\epsilon_2},
\end{equation}
from Eqs.~(\ref{PhaseEquilibrium_P})
and(\ref{PhaseEquilibrium_mu}). Hence, for the critical value we
have
\begin{equation}
\WID{\lambda}{cr}=\frac{3}{2}\left(1+\frac{P}{\epsilon_2}\right)=\frac{3}{2}\left(1+\frac{\alpha(\rho_*{-}1)}{\rho_*}\right),
\end{equation}
where the last equality is valid, naturally, only for our linear
EOS. It is this result that is indicated by the thin dotted line
in Fig.~\ref{Pix-Lambda}. Remarkably, only the quantities
referring to the central phase enter into the expression for
$\WID{\lambda}{cr}$. In addition, a condition for the parameters
of interest to us can be derived from the equilibrium equations
(\ref{PhaseEquilibrium_P}) and (\ref{PhaseEquilibrium_mu}) and the
requirement $\epsilon_1\geq 0$:
\begin{equation}
\frac{\rho_*}{\rho_*{-}1}\geq\alpha(\lambda{-}1).
\end{equation}
It bounds the range of $\rho_*$ at $\lambda>1+1/\alpha$.

\section*{DISCUSSION AND CONCLUSIONS}
 Let us briefly summarize our main results: the existence
of a special point on the mass–radius diagram of hybrid stars is a
consequence of the combined action of several factors. First, the
quark EOS for which the main local condition (\ref{MainEquation})
was shown to be met because the equilibrium equations are
homologous is linear. Second, the “phase diagram” of quark matter
has peculiarities (see Fig.~\ref{Pix-Lambda}); as a consequence,
much of the curve of stationary points lies in a small region of
the mass–radius diagram (Fig.~\ref{Pix-MP_zoom}). Finally, the
topology of the curves $M(R)$ itself favors their intersection in
a narrow region. Interesting questions arise here: First, will the
property of intersection be retained on a global scale for a
distinctly different phase diagram, i.e., at properties of the
crust differing significantly from those considered? Second, are
there solutions with other, nonlinear EOSs for our main (local)
stationarity condition (\ref{MainEquation})? And, finally, the
question touched on at the very beginning: how will our results
change for the Gibbs description of the phase transition, where a
region of mixed states appears instead of the sharp boundary
between the phases in a star? These questions need to be
investigated further.

Next, we established that the stars at the special point are
“masked”, hiding their true structure under the veil of observable
quantities ($M$ и $R$). Consider this aspect of the problem. Let
us adopt the linearity of the quark EOS and assume that we know
the true EOS of nuclear matter without any phase transitions that
gives a thick enveloping curve on the $(M{-}R)$ (see
Figs.~\ref{Pix-PureQuark} and \ref{Pix-Magic_Point_main}). Then,
were it not the special point, only one measurement of the stellar
mass and radius not only could say us whether such a star is a
purely neutron or hybrid one (or, as a limiting case, a purely
quark one) but could also point to the parameters of quark matter.
However, the existence of a special point changes the situation:
measuring the mass and radius of a star in its vicinity will only
say us that this star contains a quark core, but neither its
structure nor the parameters of quark matter will be determined.
Either invoking additional information (for example, the cooling
rate if the star was hot) or measuring the parameters of other
hybrid stars to gain statistics and reconstruct the true curve
$M(R)$ will be required.

\section*{ACKNOWLEDGMENTS}
This work was supported by grant no. 11.G34.31.0047 from the
Government of the Russian Federation and SNSF SCOPES project no.
IZ73Z0-128180/1. This work was also supported in part by the
Russian Foundation for Basic Research (project nos. 11-02-00882-a,
12-02-00955-a, and 13-02-12106). We are grateful to the anonymous
referees for their helpful critical remarks.
\pagebreak

\centerline {\bf APPENDIX}

\noindent Let us briefly describe the numerical method that we
used to find the stationary points on the mass–radius diagram. Let
we have a procedure that, starting from some central pressure
$\WID{P}{c}$ (or, alternatively $\WID{\epsilon}{c}$), integrates
the equilibrium equations (\ref{dPdr}) and (\ref{dmdr}) up to the
surface defined by the condition $\WID{P}{s}=0$. The stellar mass
and radius being obtained in this case can be written as
$M=M(\WID{P}{c},B)$ and $R=R(\WID{P}{c},B)$, where the dependence
on parameter $B$ is shown explicitly. For a small change in input
parameters, we can, naturally, write
\begin{align}
\triangle
M=&\DERS{M}{\WID{P}{c}}{B}\triangle\WID{P}{c}+\DERS{M}{B}{\WID{P}{c}}\triangle
B,\\
\triangle
R=&\DERS{R}{\WID{P}{c}}{B}\triangle\WID{P}{c}+\DERS{R}{B}{\WID{P}{c}}\triangle
B.
\end{align}
At a stationary point, the equations $\triangle M=0$ and
$\triangle R=0$ have nontrivial solutions and, hence, the
determinant of the system
\begin{equation}
\mbox{DET}\equiv\DERS{M}{\WID{P}{c}}{B}\DERS{R}{B}{\WID{P}{c}}-\DERS{M}{B}{\WID{P}{c}}\DERS{R}{\WID{P}{c}}{B}
\end{equation}
becomes zero. The derivatives in the determinant are easy to
calculate numerically using several calls of the corresponding
procedure and finite--difference equations. Thus, we obtain the
function $\mbox{DET}=\mbox{DET}(\WID{P}{c},B)$ whose zeros specify
the sought-for stationary points.

\end{document}